# Nature of the 1/$f$ Noise in Graphene – Direct Evidence for the Mobility Fluctuations Mechanism

Adil Rehman[1*], Juan Antonio Delgado Notario[1], Juan Salvador Sanchez[2], Yahya Moubarak Meziani[2], Grzegorz Cywiński[1], Wojciech Knap[1], Alexander A. Balandin[3*], Michael Levinshtein[4], Sergey Rumyantsev[1*]

[1]CENTERA Laboratories, Institute of High Pressure Physics, Polish Academy of Sciences, Warsaw 01-142, Poland

[2]Nanotechnology Group, USAL-Nanolab, Universidad de Salamanca, Salamanca 37008, Spain

[3]Nano-Device Laboratory, Department of Electrical and Computer Engineering, University of California, Riverside, California 92521, USA

[4]Ioffe Physical-Technical Institute, St. Petersburg 194021, Russia

*e-mail: adilrehhman@gmail.com; balandin@ece.ucr.edu; roumis4@gmail.com

**Abstract**

The nature of the low-frequency current fluctuations, *i.e.* carrier number *vs.* mobility, defines the strategies for noise reduction in electronic devices. While the 1/*f* noise in metals has been attributed to the electron mobility fluctuations, the direct evidence is lacking (*f* is the frequency). Here we measured noise in *h*-BN encapsulated graphene transistor under the condition of geometrical magnetoresistance to directly assess the mechanism of low-frequency electronic current fluctuations. It was found that the relative noise spectral density of the graphene resistance fluctuations depends *non-monotonically* on the magnetic field (*B*) with a minimum at approximately $\mu_0 B \cong 1$ ($\mu_0$ is the electron mobility). This observation proves unambiguously that the mobility fluctuations are the dominant mechanism of the electronic noise in high-quality graphene. Our results are important for all proposed applications of graphene in electronics and add to the fundamental understanding of the 1/*f* noise origin in any electronic device.

The history of the 1/*f* noise investigation started in 1925, almost a hundred years ago, when Johnson discovered that fluctuations of the thermionic current in a vacuum tube increase with frequency decrease[1]. Since that time, the 1/*f* noise has been discovered in an enormous number of electronic and non-electronic systems. It is conventionally accepted that the term "1/*f* noise" is applied to fluctuations with the smooth $1/f^{\gamma}$ spectrum with the slope parameter $\gamma \cong 0.8 - 1.2$. It has been soon established that the noise spectral density of the current fluctuations in semiconductor and metal resistors in the linear regime, is proportional to the voltage squared (see reviews[2-5], and references therein). This fact suggests that the 1/*f* noise is due to the resistance fluctuations, and the electrical current only reveals these fluctuations. In 1976, more than fifty years since the discovery of the 1/*f* noise, the direct proof of the resistance fluctuations as an origin of the 1/*f* noise was provided by Voss and Clark[6]. They demonstrated that the slow fluctuations of the thermal noise amplitude and the fluctuations of the current are caused by the same equilibrium fluctuations of the resistance. Moreover, the amplitude and spectrum of the resistance fluctuations were the same in both experiments.

The next important fundamental question to be answered is: *What causes these resistance fluctuations*? It can be either the fluctuations in the mobility of the charge carriers or in their number. This is not an easy question because the mobility and concentration of the charge carriers contribute symmetrically to the resistance $R \sim 1/N\mu$ ($N$ is the number of the charge carriers and $\mu$ is the mobility). The contribution of the mobility and of the number of carrier's fluctuations can be different depending on the devices and materials. Despite numerous attempts to answer this fundamental question, there is still no conventionally accepted point of view on this problem. The McWhorter model describes accurately the 1/*f* noise in many semiconductor devices by attributing it to the fluctuations in the number of charge carriers[7]. While many studies attribute noise in metals or certain electronic devices to the mobility fluctuations, the evidences for such fluctuations as the true source of 1/$f$ noise are more controversial.

**1/$f$ noise in conventional semiconductors**

Theory of the electron transport in magnetic field provides an elegant way for testing the nature of the 1/$f$ noise. In 1983, some of us have used noise measurements in magnetic field in order to distinguish between the mobility ($\delta\mu$) and the number of carriers ($\delta N$) fluctuations[8]. In the sample geometry with the length (*L*) much smaller than its width (*W*), i.e., *L*<<*W*, or in the Corbino disc geometry, the magnetic field perpendicular to the electrical field causes the well-known effect of the geometrical magnetoresistance[9]

$$\rho_{xx} = \rho_0[(1 + (\xi\mu_0 B)^2]. \tag{1}$$

Here $\rho_{xx}$ is the longitudinal resistivity, $\mu_0$ is the mobility at magnetic field *B*= 0 T, and $\xi$ is the scattering factor, which depends on the details of the scattering mechanisms, *e.g.* in semiconductors, $\xi \geq 1$. Differentiating Eq. (1), one can show that the mobility fluctuations do not contribute to the resistance fluctuations when the condition $\xi\mu_0 B = 1$ is satisfied:

$$\frac{\delta\rho_{xx}}{\rho_{xx}} = \frac{\delta\mu_0}{\mu_0} \frac{(\xi\mu_0 B)^2 - 1}{(\xi\mu_0 B)^2 + 1}. \tag{2}$$

Therefore, if the 1/$f$ noise is caused by the mobility fluctuations one can expect a strong decrease of the noise at $\mu_0 B \cong 1$. On the other hand, the concentration and the total number of carriers do not depend on the magnetic field, and the noise, in the case of the dominance of the fluctuations in the number of carriers, should not depend on the magnetic field.

The measurements of the 1/$f$ noise in a semiconductor such as GaAs, under the condition of the strong geometrical magnetoresistance, showed that the noise does not depend on the magnetic field[8]. The latter led to the conclusion that the fluctuations in the number of charge carriers were the source of the 1/$f$ noise. It was also shown that the noise was not originating on the surface of the semiconductor but rather in its volume. Similar experiments with the same conclusions have been repeated by other groups for GaAs, and by some of us for GaN[10-14]. Other investigations of the nature of noise in semiconductors also complied with the number of carriers' fluctuations as the source of the 1/$f$ noise. Among the alternative experimental evidence, one can indicate the gate-voltage dependence of noise in the field-effect transistors (FETs), especially in $n$-channel Si metal-oxide-semiconductor field-effect transistors (MOSFETs). It is conventionally accepted that, with rare exceptions, the 1/$f$ noise in MOSFETs complies with the McWhorter model[7], which assumes the fluctuations in the number of carriers. This model predicts, and the experiments confirm, in the majority of cases, that when the concentration($n_s$), in the channel is tuned by the gate voltage, the noise depends on concentration as 1/$n_s^2$. Another notable proof of the carrier-number fluctuations demonstrated that individual defects in microstructures cause random telegraph signal (RTS) noise due to the charge carriers trapping and de-trapping[15]. The level of this type of noise differed from a device to a device due to the statistical variance in the number and the properties of the defects. Averaging of the noise spectra from several dozen devices yielded the same 1/$f$ noise spectrum as the one measured in larger devices obtained via the same fabrication process. These results allowed the authors to conclude that "There can now be no doubt that measurements of noise and RTSs in microstructures have shown definitively that l/$f$ noise in MOSFETs and MIM tunnel diodes is generated through carrier trapping"[15].

**Prior attempts to prove 1/*f* mobility noise in graphene**

While numerous trusted experiments demonstrate that the 1/*f* noise in semiconductors is of the number of carriers' fluctuation nature, there is no direct and conclusive evidence of the mobility fluctuations as the source of the 1/*f* noise in any material system. From the other side, it is generally accepted now that the physical origin of the 1/*f* noise is not universal and may differ from one material system to another (see for example[4]). Therefore, it is possible that noise in some materials and devices can be of the mobility fluctuations origin. There have been attempts by some of us for reproducing the experiments of noise measurements under the condition of strong geometrical magnetoresistance for the materials where mobility fluctuations may dominate the 1/*f* noise. For a long time, the best candidates for such materials were considered to be metals[4]. However, it is difficult to satisfy the condition $\mu B = 1$ in metals for a reasonable magnetic field owing to the low electron mobility in metals at room temperature (RT). Cooling down the metals increases the electron mobility but also decreases the resistivity, making the noise measurements extremely challenging, especially for the samples with the *W>>L* geometry. Some of us have made several attempts of measuring noise in metals at low temperatures in a strong magnetic field but failed to obtain conclusive results.

In 2004, a new opportunity emerged for proving directly a possibility of the mobility-fluctuation nature of the 1/*f* noise. The latter came with the exfoliation of graphene, characterized by many unique properties[16-18]. Among them are the high charge carriers' mobility at RT and unusual 1/*f* noise behavior. A large number of reports devoted to the low-frequency noise in graphene have been published (see ref. [19] and references therein). The electronic noise was typically studied in the back-gated or top-gated graphene FETs. It was found that the gate-voltage dependence of noise can be different in different graphene devices, especially close to the charge neutrality point, described by the "V", "M" and "Λ" shapes of the noise spectral density dependence on the gate-voltage. The important conclusion from all studies was that the overall gate-voltage dependence of noise in graphene FETs is weak, and it does not comply with the McWhorter model. This observation led to the suggestion that mobility fluctuations can possibly constitute the physical mechanism of the 1/$f$ noise in graphene[20].

A scenario when the mobility fluctuations are responsible for the 1/$f$ noise can be envisioned theoretically. The latter can be a result of the scattering cross-section fluctuations. In the case of only one type of the fluctuating scattering center, the noise spectral density of the resistance fluctuations is given as[21,22]

$$\frac{S_R}{R^2} = \frac{4N_{t\mu}\tau P(1-P)}{1+(\omega\tau)^2} l_0^2 (\sigma_2 - \sigma_1)^2 \qquad (3)$$

where $N_{t\mu}$ is the concentration of the scattering centers of a given type, $l_0$ is the mean free path of the carriers, $P$ is the probability for a scattering center to be in the state with the cross-section $\sigma_1$, and $\tau$ is the characteristic time constant. There can be different reasons for fluctuations of the scattering cross-section, *e.g.*, change of the traps' charge state, motion of the dislocations and individual atoms, or diffusion of defects. Since these processes can have different values of $\tau$, $\sigma$, and $P$, the integration over these parameters leads to the 1/$f$ or 1/$f$-like noise spectrum. One should note that the relative noise spectral density in Eq. (3) does not depend either on the free carrier concentration or on the total number of carriers in the sample. This complies with the weak gate-voltage dependence observed for the low-frequency noise in graphene FETs[23]. Another indirect indication that the 1/$f$ noise in graphene can be of the mobility fluctuation origin was obtained in the experiments with graphene FETs irradiated with the electron beam[24]. As expected, the irradiation resulted in the introduction of additional defects and in the reduction of the electron mobility. Surprisingly, irradiation led also to a decrease in the 1/$f$ noise level. This unusual behavior of the 1/$f$ noise in graphene can be explained if one assumes the mobility fluctuation mechanism and adopts Eq. (3). Indeed, the radiation-induced introduction of defects leads to the decrease in the mobility and mean free path, with the corresponding decrease in noise. One should note here that the defects, which mostly limit the electron mobility and the trapping centers predominantly responsible for the 1/$f$ noise are not necessarily the same. Similar results were obtained by other independent research groups using different types of irradiation[25,26].

The prior studies of noise in graphene suggested that this material is an excellent candidate for demonstrating directly that the mobility fluctuations can be the mechanism of the 1/$f$ noise via experiments with the magnetic field. In 2013, some of us made the first attempt at such study[27]. The reduction of the 1/$f$ noise in graphene FETs subjected to a magnetic field was observed at cryogenic and near RT temperatures. However, the noise dependence on the magnetic field strength was different from that predicted by Eq. (1). The studied graphene samples were also characterized by strong physical rather than geometrical magnetoresistance. The possible reason for that could be the high concentration of the defects in the studied samples and a rather low charge carrier mobility. Therefore, the conclusion of the nature of the 1/$f$ noise could not have been made at that time. To sum up, the multiple prior efforts at demonstrating that the mobility fluctuations can be the dominant source of the 1/$f$ noise in any electronic material system have failed. One can only state that there have been experiments suggesting indirectly that 1/$f$ noise can be of the mobility fluctuation type in graphene. These are the experiments with the gate-bias dependence and irradiation effects on noise mentioned above.

**Geometrical magnetoresistance experiments graphene FETs**

For this study, we used high-quality single layer graphene encapsulated in $h$-BN to fabricate back-gated FETs. The device design and high-quality graphene allowed us to achieve the electron mobility of ~3 m$^2$/Vs at room temperature. It is important that such high mobility value was achieved in the graphene FETs on substrates, rather than suspended graphene, to ensure proper gating and Fermi level tuning. The measurements were intentionally performed at relatively high temperatures of 200 K - 300 K in order to avoid complications in the data interpretation due to quantum confinement effects possible at low temperatures. The aspect ratio of the FET rectangular channel was chosen to be relatively small (*W/L*=4). In comparison with a higher *W/L* ratio, these specific dimensions allowed us to minimize the contribution of the contact resistance. Figures 1a and 1b show the schematic of the *h*-BN encapsulated transistor and optical microscope image of two representative devices, respectively. The current-voltage (*I-V*) characteristics and the low-frequency noise were measured inside the closed cycle cryogenic probe station using conventional instrumentation (Lake Shore). Figure 1c presents the transfer *I-V* characteristics of the graphene FET measured at two temperatures. It is seen that the Dirac charge neutrality voltage ($V_D$), does not depend on temperature. Away from the Dirac point, there is a region of the bias voltage where the *I-V* characteristics are linear. The *I-Vs* start to saturate at higher voltages. There are several possible reasons for this saturation[28,29]. One of them, most relevant in the context of the present study, is the contact resistance ($R_c$), which may become comparable with the channel resistance at high carrier's concentration.

The upper bound of the contact resistance can be estimated assuming the concentration independent mobility. In this case, the extrapolation of the resistance versus $(V_D-V_G)^{-1}$ dependence to $(V_D-V_G)^{-1}=0$ yields a rough estimate of the total contact resistance. Using this procedure, we obtained $R_c$ =100 Ω and 140 Ω for the temperatures 300 K and 200 K, respectively. The noise was measured at *V*g=0. This bias point corresponds to the linear part of the *I-V* characteristic that ensures the small effect of the contact resistance. Based on the location of this bias voltage with respect to the Dirac point, the electrons dominated the electrical conductivity and noise properties.

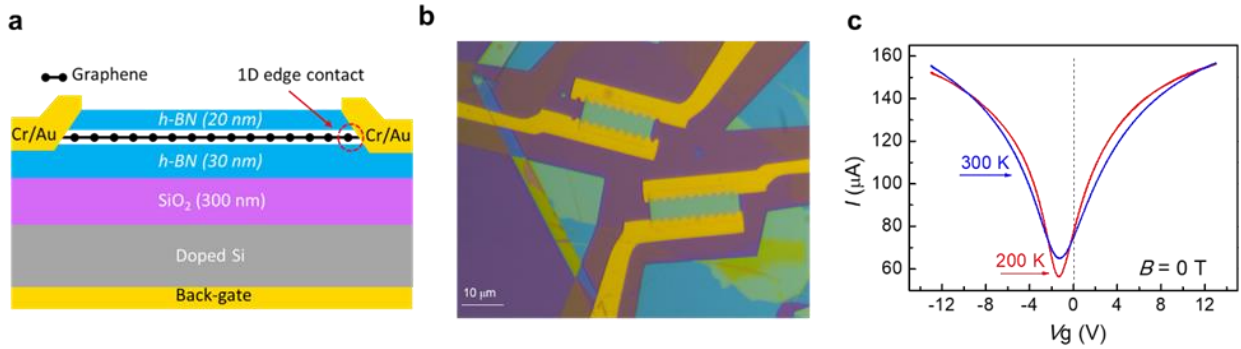

**Fig. 1 | Graphene device design and current-voltage characteristics. a**, Schematic and **b**, optical microscopy image of *h*-BN encapsulated graphene FETs used in the study. **c**, Transfer current-voltage characteristic of the representative graphene FET. The *I-Vs* are shown for 300 K and 200 K. The vertical dashed line indicates the bias voltage used for the noise measurements.

It is known that in graphene the magnetoresistance can show a complicated behavior with the magnetic field ($B$) with the dependencies ranging from $\sqrt{B}$ to $B^2$. The type of dependence is defined by the specific scattering mechanisms, which limit the electron transport[30,31]. Recent detailed measurements of the magnetoresistance in Corbino disks fabricated from high quality suspended graphene demonstrated quadratic dependence of the magnetoresistance[31]. These results confirm the validity of Eq. (1), at least for the specific scattering mechanisms. The results of our magnetoresistance measurements are presented in Fig. 2a,b. To analyze the results, we take into account the contact resistance and a finite *W*/*L* ratio:

$$R(B) = R_{\text{Ch}}[1 + (\eta\xi\mu_0 B)^2] + R_\text{c}. \tag{4}$$

Here $R_{Ch}$ is the resistance of the channel without a magnetic field ($B=0$ T), $R_c$ is the total contact resistance, and $\eta$ is the geometrical factor. In small magnetic fields $\mu_0 B<1$, the geometrical factor can be approximated as $\eta^2=(1-0.54L/W)$ [32]. One can see in Fig. 2a, that in a weak magnetic field, the relative magnetoresistance, indeed, depends quadratically on the magnetic field, *i.e.* $[R(B)-R(0)]/R(0) \sim B^2$. In high magnetic fields, this dependence becomes weaker (see Fig. 2b) so that it can be approximated as $[R(B)-R(0)]/R(0) \sim B$. There are several reasons for the deviation from the quadratic magnetoresistance, which include the dependences of the momentum relaxation time and the cyclotron frequency on the magnetic field, as well as the influence of the sample borders. The analysis of the magnetoresistance in a high magnetic field in graphene is beyond the scope of the present work.

We use Eq. (4) to estimate the *apparent* electron mobility value, corrected to $\eta\xi$ factor:

$$\eta\xi\mu_{app} = \left(\frac{R(B)-R(0)}{R_{Ch}+R_c}\frac{1}{B^2}\right)^{0.5} \tag{5}$$

From the measured data, we obtain $\eta\xi\mu_{app} \cong 3$ m$^2$/Vs at 300 K and $\eta\xi\mu_{app} \cong 3.8$ m$^2$/Vs at 200 K. The actual electron mobility relates to the apparent mobility as

$$\mu_0^2 = \mu_{app}^2 \frac{R_{Ch}+R_c}{R_{Ch}}. \tag{6}$$

The resistance fluctuations owing to the mobility fluctuation can be obtained by differentiating Eq. (4), resulting in the expression:

$$\frac{\delta R(B)}{R(B)} = \frac{\delta\mu_0}{\mu_0}\left[\frac{(\eta\xi\mu_0 B)^2-1}{(\eta\xi\mu_0 B)^2+1+R_c/R_{Ch}}\right] \tag{7}$$

The noise spectral density, corresponding to these resistance fluctuations, can be written in the following form:

$$\frac{S_R(B)}{R(B)^2} = \left[\left(\frac{R_{Ch}}{R_{Ch}+R_c}\right)^2 \frac{S_{\mu 0}}{\mu_0^2}\right] \left[\frac{(\eta\xi\mu_{app}B)^2 \frac{R_{Ch}+R_c}{R_{Ch}} - 1}{(\eta\xi\mu_{app}B)^2 + 1}\right] + C \qquad (8)$$

Here, we added a constant (*C*) which represents residual noise remaining when the noise due to the mobility fluctuations is suppressed. The first factor in Eq. (8) is the noise at zero magnetic field. The parameter $\eta\xi\mu_{app}$ is known from the magnetoresistance measurements. Therefore, the fitting parameters in Eq. (8) are the contact resistance ($R_c$), and the residual noise level (*C*). The upper bound of the contact resistance is known from the I-V characteristics.

The low-frequency noise spectra measured at 300 K and 200 K were always of $1/f^\gamma$ type with $\gamma \sim 1$. Figure 2c shows a few examples of the noise spectra measured at 200 K and in different magnetic fields. The noise spectral density of the resistance fluctuations only weakly depended on the gate voltage and was independent of the drain voltage.

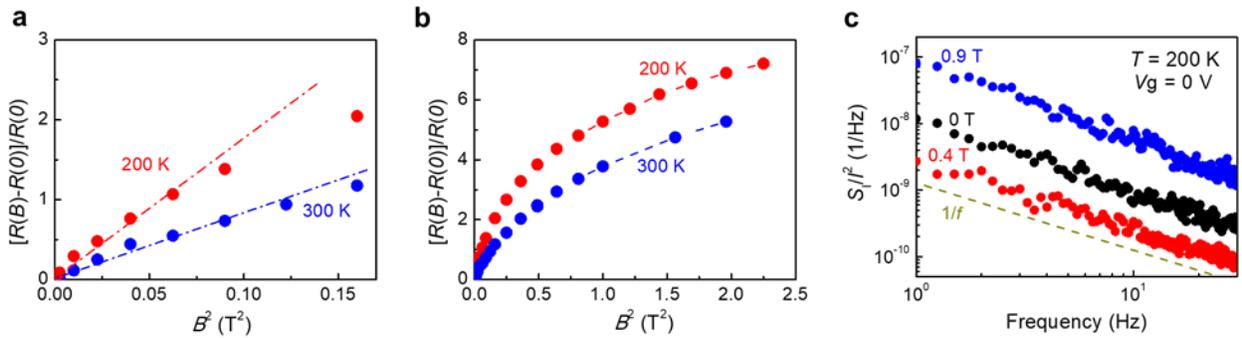

**Fig. 2 | Relative magnetoresistance as a function of the magnetic field and examples of the noise spectra.** The [*R(B)-R(0)/R(0)*] dependence is shown in the **a,** low **b,** and high magnetic fields

at the temperatures of 300 K and 200 K. **c**, Noise spectra are shown at 200 K at different magnetic fields.

The dependences of noise spectral density on the magnetic field are shown in Fig. 3 for T = 300K and T = 200K. The symbols are the experimental dependence of noise at the frequency of the analysis $f$=10 Hz. The data are shown for two temperatures. The continuous solid and dashed lines represent the results of the calculations using Eq. (8). The solid lines correspond to $R_c$=0 while the dashed lines are obtained for the $R_c$ values extracted from the I-V characteristics. The contact resistance represents the upper bound, and it is most likely overestimated. Therefore, the actual dependence of noise on the magnetic field falls into the interval between the solid and dashed lines.

The first important observation from Fig. 3 is that Eq. (8) represents well the decrease of the noise in weak magnetic fields. In excellent agreement with the assumption of the mobility fluctuations as an origin of the 1/$f$ noise, the magnetic-field dependence of noise has a pronounced minimum at the field, which corresponds approximately to the condition $\mu_0 B$=1. Note, that we should not expect the perfect match of the experiment and calculations for the position of the noise minimum because, as one can see from Fig. 2b the magnetoresistance deviates from the quadratic dependence at high magnetic fields. Although Eq. (8) reproduces well the overall magnetic field dependence of noise, the higher is the magnetic field, the less is an agreement between the experimental data and the model. This is predictable because in high magnetic fields the magnetoresistance strongly deviates from $[R(B)-R(0)]/R(0) \sim B^2$ law. There are several reasons for the strong increase of noise in high magnetic fields. First, in the high magnetic field, electrons are pushed to one of the side edges of the channel. Edge states might give a higher contribution to noise. Second, the current crowding at the rectangular sample corners gives an enhanced contribution to noise.

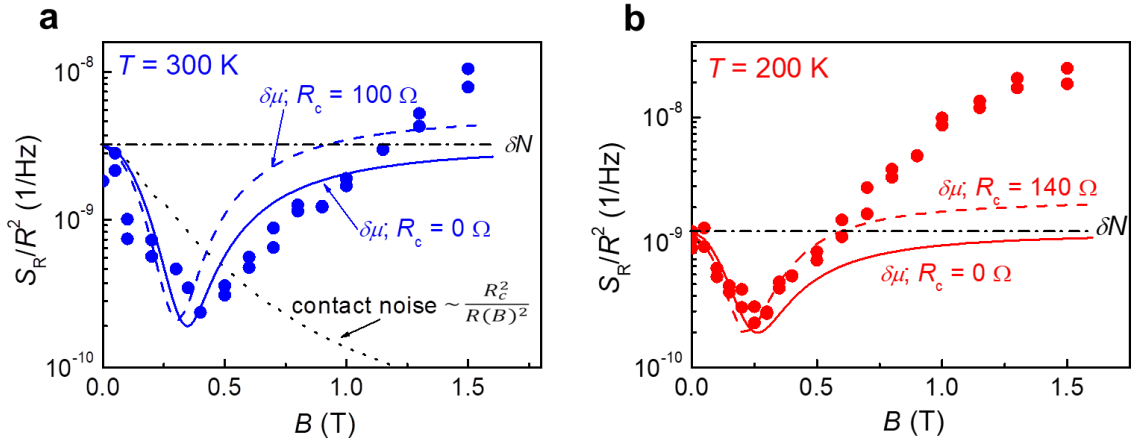

**Fig. 3 | Relative noise spectral density as a function of the magnetic field.** The results are shown for graphene at **a**, $T = 300$ K and **b**, $T = 200$ K under the condition of the geometrical magnetoresistance in FETs with the channel's aspect ratio $W/L = 4$. The symbols show the experimental results for the frequency of analysis $f = 10$ Hz. The solid and dashed lines are the calculation assuming the mobility fluctuations as a dominant source of the $1/f$ noise. The solid line corresponds to the $R_c = 0$ Ω. The dashed line corresponds to the upper bound of the contact resistance extracted from current-voltage characteristics. The dotted and dash-dotted lines show the hypothetical noise behavior in the cases of the contact noise and noise due to the number of carrier's fluctuations, respectively.

To add the final element to the conclusion about the noise nature, we have to exclude a possible contact origin of the noise. If noise originates from the contact resistance fluctuations, the noise spectral density of the total resistance fluctuations can be obtained from the analysis of the channel and contact resistors series connection

$$\frac{S_R(B)}{R(B)^2} = \frac{S_{Rc}}{R_c^2} \frac{R_c^2}{R(B)^2}, \tag{9}$$

where $S_{Rc}/R_c^2$ is the relative spectral noise density of the contact resistance fluctuations, which does not depend on the magnetic field. In Fig. 3a, the dotted line shows the dependence calculated using Eq. (9) and experimental dependence $R(B)$. As seen the assumption of the contact resistance noise does not agree with the experiment, either in weak or strong magnetic fields.

**Conclusions**

For the first time since the discovery of the 1/$f$ noise hundred years ago, we have proven conclusively that the fluctuations in the mobility of charge carriers can be the dominant mechanism of the 1/$f$ noise. We utilized the unique geometry and high electron mobility of graphene to directly assess the mechanism of the low-frequency electronic current fluctuations. In order to do this, the noise was measured as a function of the magnetic field strength under the condition of geometrical magnetoresistance. It was found that the relative noise spectral density of the graphene resistance fluctuations depends non-monotonically on the magnetic field with a minimum at $\mu_0 B \cong 1$. This observation proves unambiguously that the mobility fluctuations are the dominant mechanism of the low frequency noise in graphene. Our results are important for all proposed applications of graphene in electronics, since the 1/$f$ noise is the dominant contributor to the phase noise in electronic communications systems, and limits the sensitivity and selectivity of the electronic sensors. The significance of our finding goes beyond the graphene field by adding to the fundamental understanding of 1/$f$ noise: mobility fluctuations like carrier number fluctuations can indeed have the relevant time constants for dominating the noise spectrum at low frequencies.

**Methods**

**Heterostructure Fabrication and Characterization:** Graphene field-effect transistors were fabricated from the graphene-based heterostructures. Fabrication of the graphene heterostructures started from the mechanical exfoliation of monolayer graphene and two relatively thick hexagonal boron nitride (*h*-BN) flakes on a SiO$_2$/Si substrate with 300 nm SiO$_2$. To achieve a larger size of the graphene flake, the Si/SiO$_2$ substrate was exposed to an oxygen plasma (50 W, 120 s), and thereupon the surface of the tape was brought into contact with the SiO$_2$ surface. As the next step, the substrate was annealed on the hot plate for 2 mins at 100 °C, and after cooling down the sample to room temperature the tape was removed. Finally, the substrate was examined under the optical microscope to identify the graphene flakes (Fig. S1a). Oxygen plasma pre-treatment on the substrates was not necessary for the exfoliation of the *h*-BN flakes as we were able to find enough large *h*-BN flakes (Fig. S1b,c). Both *h*-BN flakes were characterized by a Stylus Profilometers. We found a thickness of ~20 nm for the top *h*-BN and ~30 nm for the bottom *h*-BN.

The graphene-based heterostructure was fabricated by using a polymer dry-transfer technique, similar to the method for assembling van der Walls heterostructures reported in ref. [33]. The stamp to pick-up and transfer the required 2D flakes was prepared by using a homemade polycarbonate (PC) solution (6.5 % dissolved in chloroform) and a polydimethylsiloxane (PDMS) block. The staking process starts as follows: the top *h*-BN flake was picked up and then transferred onto the graphene sheet to initially fabricate a *h*-BN/graphene heterostructure. After that, the *h*-BN/graphene heterostructure was cleaned in chloroform for a few minutes to remove the PC residues. The process was repeated, and the heterostructure was picked up and transferred onto the bottom h-BN to fabricate the final *h*-BN/graphene/*h*-BN van der Waals heterostructure (see Fig. S2a). The resulting van der Waals heterostructure was characterized by Raman spectroscopy (micro-Raman spectrometer LabRAM HR Evolution) using a 532 nm laser with the incident power of approximately 1 mW. Typical Raman spectrum is shown in the Fig. S2b.

**Devices Fabrication:** The device fabrication process followed the procedure described in details in ref. [34,35]. The fabricated heterostructure was first patterned using the electron beam lithography (EBL) and a homemade PMMA (5% in chlorobenzene) as a resist to define two bars with widths, W, of 16 µm and 20 µm (Fig. S3a). The heterostructure was dry-etched for 20 seconds in an ICP-RIE Plasma Pro Cobra 100 with a SF6 atmosphere (40 sccm, P = 75 W and T = 10 ºC), and the areas uncovered by a PMMA, used as a mask, were removed. Then, the graphene based heterostructure bars were cleaned in acetone and isopropanol for a few minutes to remove the remaining PMMA and checked in the optical microscope (Fig. S3b). The drain and source contacts were patterned by a second round of e-beam lithography (Fig. S4a), and thereupon the sample was dry-etched once again to open accesses to the encapsulated graphene sheet[2]. Our etching recipe provides a truncated square pyramid shape with a contact angle of approximately 40º to the horizontal plane ensuring high-quality quasi one-dimensional ohmic contacts with the encapsulated graphene. Finally, the sample was placed inside the e-beam evaporator where the metal drain and source contacts were made by evaporating at very low pressures ($10^{-8}$ mbar) 3.5 nm of Cr (0.06 nm/s) and 55 nm of Au (0.25 nm/s). The final device, after a lift-off process in acetone, is shown in Fig. S4b. The drain and source contacts were patterned with a meander shape in order to increase the contact length and minimize the contact resistance.

**Current-voltage and noise measurements:** The current-voltage characteristics and noise spectra were measured inside a closed cycle cryogenic probe station (Lake Shore Inc. CRX-VF). The noise was measured with the source and gate grounded. The signal from the drain load resistor ($R_L$) was amplified and analyzed by the Photon Spectrum Analyzer. To minimize the power supplier noise, we used a battery biasing circuit to apply voltage bias to the devices. The noise spectral density of the drain voltage fluctuations ($S_V$) was converted to the current ($S_I/I^2$), and resistance ($S_R/R^2$) relative noise spectral densities as:

$$\frac{S_I}{I^2} = \frac{S_R}{R^2} = S_V \left(\frac{R_L + R}{R_L R}\right)^2 \frac{1}{I^2}$$

**Data availability:** The data that support the plots within this paper and other findings of this study are available from the corresponding author upon reasonable request.

**ACKNOWLEDGMENTS**

Authors are very grateful to Prof. M.I.Dyakonov, who proposed the idea of the noise measurements under magnetoresistance conditions in 1982. The work was supported by the CENTERA Laboratories in the framework of the International Research Agendas program for the Foundation for Polish Sciences, co-financed by the European Union under the European Regional Development Fund (No. MAB/2018/9). A.A.B. acknowledges the Vannevar Bush Faculty Fellowship award (class of 2021). M.M. acknowledges the support by the Spanish Ministry of Science, Innovation, and Universities and FEDER under the Research Grants numbers RTI2018-097180-B-100, and FEDER/Junta de Castilla y León Research Grant numbers SA256P18 and SA121P20. J.A.D.N also acknowledges the technical help from Dr. Vito Clericò.

**Author Contributions**

M.L. and S.R conceived the idea of the noise measurements in graphene in magnetic field. A.R. and S.R. planned and performed the measurements. J.A.D.N., J.S.S., and M.M. fabricated and characterized the samples. G.C. W.K., and A.A.B. contributed to the noise data analysis. All authors participated in discussions, analysis and manuscript preparation.

**Competing interests**

The authors declare no competing interests.

**Correspondence and requests for materials** should be addressed to A.R., A.A.B., S.R.